\documentclass[final,3p,times,twocolumn]{elsarticle}
 \biboptions{comma,sort&compress}
\usepackage{here}
 \usepackage{graphicx}
 \usepackage{epsfig}
\usepackage{epstopdf}
\usepackage{amsmath}

\def\nin{\noindent}
\def\beq{\begin{equation}}
\def\eeq{\end{equation}}
\def\bea{\begin{eqnarray}}
\def\eea{\end{eqnarray}}
\def\nnb{\nonumber}
\def\la{\langle}
\def\ra{\rangle}
\def\ga{\left(}
\def\dr{\right)}
\def\lrar              {\Longrightarrow}

\journal{Nuc. Phys. (Proc. Suppl.)}

\begin{document}

\begin{frontmatter}



\title{$ \bar D^*D$ and $\bar B^*B~(1^{++})$ molecules at N2LO from QSSR$^*$}


 \author[label1]{F. Fanomezana}
  \address[label1]{Institute of High Energy Physics of Madagascar (iHEP-MAD), University of Antananarivo, Madagascar }

\ead{fanfenos@yahoo.fr}

 \author[label2]{S. Narison}
  \address[label2]{Laboratoire
Univers et Particules (LUPM), CNRS-IN2P3 \& Universit\'e
de Montpellier II, 
\\
Case 070, Place Eug\`ene
Bataillon, 34095 - Montpellier Cedex 05, France.}
\ead{snarison@yahoo.fr}

 \author[label1]{A. Rabemananjara\corref{cor2}}
\cortext[cor1]{Talk given at  QCD 14 (29 june - 3 july, Montpellier - France).}
\cortext[cor2]{Speaker.}
\ead{{achris$\_$}01@yahoo.fr.}


\begin{abstract}
\noindent
We use QCD spectral sum rules (QSSR) and the factorization properties of molecule currents to estimate  the masses and couplings  of the $\bar D^*D$ and $\bar B^*B~(1^{++})$ molecules at N2LO of PT QCD. We include in the OPE the contributions of non-perturbative condensates up to dimension-eight. With the Laplace sum rules approach (LSR) and in the $\overline{MS}$-scheme, we obtain $M_{D^*D}=3738(152)$ MeV, which agrees within the errors with the newly discovered $Z_c$(3900). For the bottom channel, we find   $M_{B^*B}=10687(232)$ MeV in good agreement with the observed $Z_b$(10610). Couplings of these states to the currents are also extracted. Our results are improvements of the LO ones in the existing literature.

\end{abstract}

\begin{keyword}
QCD Spectral Sum Rules, molecule states, heavy quarkonia.


\end{keyword}

\end{frontmatter}


\vspace*{-1cm}
\section{Introduction}
\vspace*{-0.2cm}
\nin
 The recent discovery of the $Z_c$(3900) $1^{++}$ by Belle \cite{BELLE} and BESIII \cite {BES} from its $J/\psi\pi^\pm$ decays has motivated different theoretical analysis\,\cite{REVMOL}.
  However, all of the previous analysis like e.g. in \cite{RAPHAEL} from QCD Spectral Sum Rules (QSSR) \cite{SVZ,SNB} have been done at LO of PT QCD. In this paper, we are going to use QSSR to evaluate the mass and coupling of the $1^{++}$ $\bar D^*D$ and $\bar B^*B$ molecules at N2LO in the PT series and compare the results with those obtained at lowest order and with experiments. 
\vspace*{-0.25cm}
\section{QCD analysis of spin one molecule}
\vspace*{-0.25cm}
\nin
   \subsubsection*{$\bullet$ Current and two-point fonction}
   \nin
The current for this molecule state is given by:
\bea
 J^\mu &\equiv&(\bar Q\gamma^\mu q)(\bar q \gamma^5 Q )~,\\
 \nonumber Q &\equiv&  c, b\ ~~ {\rm and}~~ \ q \equiv u, d~.
 \eea
The associated  two-point correlation function is:
\bea
\Pi^{\mu\nu}_{mol}(q)&=&i\int d^4x ~e^{iq.x}\la 0
|TJ^\mu(x){J^\nu}^\dagger(0)
|0\ra\nnb\\
&=&-(q^2 g^{\mu\nu}-{q^\mu q^\nu})\Pi_{mol}(q^2)\nnb\\
&&+q^\mu q^\nu\Pi_{mol}^{(0)}(q^2)~,
\label{2po}
\eea
where $\Pi_{mol}$ and  $\Pi_{mol}^{(0)}$ are respectively associated to the spin 1 and 0 molecule states. 
In the QSSR method and parametrizing the spectral function by one resonance plus a QCD continuum, the lowest resonance mass $M_H$ and coupling $f_H$ normalized as:
\bea
\la 0|J^\mu|H\ra=f_H M_H^4 \epsilon^\mu~,
\label{eq:decay}
\eea
can be extracted by using the Laplace sum rules (LSR) which gives two well-known sum rules\,\cite{SNB}:
\beq
M^2_H=\frac{\int_{4m^2_Q}^{t_c} dt~ t ~ e^{-t \tau}  \frac{1} {\pi} {\rm Im} \Pi^{OPE}_{mol}(t)}{\int_{4m^2_Q}^{t_c} dt~ e^{-t \tau}  \frac{1} {\pi} {\rm Im} \Pi^{OPE}_{mol}(t)}
\label{mass}
\eeq
and
\beq
f^2_H=\frac{\int_{4m^2_Q}^{t_c} dt~ e^{-t \tau}  \frac{1} {\pi} {\rm Im} \Pi^{OPE}_{mol}(t)}{e^{-\tau M^2_H}M^8_H}
\label{coupling}
\eeq
where $m_Q$ is the heavy quark mass, $\tau$ the sum rule parameter and $t_c$ the continuum threshold.
\vspace*{-0.1cm}
 \subsubsection*{$\bullet$ The QCD two-point function at N2LO}
 \nin
To derive the results at N2LO, we assume factorization and then use the fact that the two-point function of molecule state can be written as a convolution of the spectral functions associated
to quark bilinear currents. In the spin one case, we have \cite{PICH,SNPIVO}:
\bea
&&\frac{1}{\pi}{\rm Im}\Pi_{mol}^{(1)}(t)=\theta(t-4M^2_Q)\left(\frac{1}{4\pi}\right)^2 t^2 \int^{(\sqrt{t}-M_Q)^2}_{M_Q^2}\hspace*{-1cm} dt_1\times\nonumber\\ 
&& \int^{(\sqrt{t}-\sqrt{t_1})^2}_{M_Q^2} \hspace*{-1cm} dt_2 ~ \lambda^{3/2} \frac{1}{\pi}{\rm Im}\Pi^{(1)}(t_1) \frac{1}{\pi}{\rm Im}\Pi^{(0)}(t_2) 
\label{conv}
\eea   
with the phase space factor:
\bea
\lambda=\left(1-\frac{(\sqrt{t}-\sqrt{t_1})^2}{t}\right)\left(1-\frac{(\sqrt{t_1}+\sqrt{t_2})^2}{t}\right)~.
\eea   
 Im$\Pi^{(1)}(t)$ and Im$\Pi^{(0)}(t)$ are respectively the spectral function associated to the vector and   to the pseudoscalar bilinear currents.
The QCD expression of the spectral functions for bilinear currents are already known up to order $\alpha_s^2$ and including non-perturbative condensates up to dimension 6. It can be found in \cite{SNFB12,SNFBST14,GENERALIS,CHET} for the on-shell mass $M_Q$.
We shall use the relation between the on-shell $M_Q$ and the running mass $\bar m_Q(\nu)$ to transform the spectral function into the $\overline{MS}$-scheme \cite{SPEC1,SPEC2}: 
\bea
M_Q &=& \overline{m}_Q(\nu)\Bigg{[}
1+{4\over 3} a_s+ (16.2163 -1.0414 n_l)a_s^2\nnb\\
&&+\ln{\ga\nu\over M_Q\dr^2} \ga a_s+(8.8472 -0.3611 n_l) a_s^2\dr\nnb\\
&&+\ln^2{\ga\nu\over M_Q\dr^2} \ga 1.7917 -0.0833 n_l\dr a_s^2\Bigg{]},
\label{eq:pole}
\eea
where $n_l=n_f-1$ is the number of light flavours and $a_s(\nu)=\alpha_s(\nu)/\pi$ at the scale $\nu$. 
\subsubsection*{$\bullet$ QCD parameters}
\nin
The PT QCD parameters which appear in this analysis are $\alpha_s$, the charm and bottom quark masses $m_{c,b}$ (the light quark masses have been neglected). 
We also consider non-perturbative condensates from \cite{BC} up to dimension 8 which are the quark condensate $\langle\bar q q\rangle$, the two-gluon condensate $\langle g^2 G^2\rangle$, the mixed condensate $\langle g\bar q G q\rangle$, the four-quark condensate $\rho\langle\bar q q\rangle^2$, the three-gluon condensate $\langle g^3 G^3\rangle$, and the two-quark multiply two-gluon condensate $\rho\langle\bar q q\rangle\langle g^2 G^2\rangle$
  where $\rho$ indicates the deviation from the four-quark vacuum saturation. Their values are given in Table \ref{tab:parameter}. For the condensates, we shall use these expressions: 
\bea
 \langle\bar q q\rangle (\nu)=-\hat\mu^3_q\left(\rm{Log}\frac{\nu}{\Lambda}\right)^{-2/{\beta_1}}\\
\langle g\bar q G q\rangle(\nu)=-M^2_0\hat\mu^3_q\left(\rm{Log}\frac{\nu}{\Lambda}\right)^{-1/{3\beta_1}}
\eea   
where $\beta_1=-(1/2)(11-2n_f/3)$ is the first coefficient of the $\beta$ function, $\hat \mu_q$ the renormalization group invariant condensate and $\Lambda$ is the QCD scale.
\vspace*{-0.25cm}
\begin{table}[hbt]
\setlength{\tabcolsep}{1.2pc}
     {\small
\begin{tabular}{ll}
&\\
\hline
Parameters&Values.    \\
\hline
$\alpha_s(M_\tau)$&0.325(8)\\
$\Lambda(n_f=4)$& $(324\pm 15)$ MeV \\
$\Lambda(n_f=5)$& $(194\pm 10)$ MeV \\
$\bar m_c(m_c)$&$(1261\pm 24)$ MeV\\
$\bar m_b(m_b)$&$(4177\pm 22)$ MeV\\
$\hat \mu_q$&$(263\pm 7)$ MeV\\
$M_0^2$&$(0.8\pm 0.2)$ GeV$^2$\\
 $\langle\alpha_s G^2\rangle$&$(7\pm 2)\times 10^{-2}$ GeV$^4$\\
 $\langle g^3 G^3\rangle$&$(8.2\pm 2.0)$ GeV$^2 \times \langle \alpha_s G^2\rangle $\\
$\rho=\langle\bar q q\bar qq\rangle/\langle\bar q q\rangle^2$&$(2\pm 1)$ \\
\hline

\end{tabular}
}
\caption{\scriptsize    QCD input parameters (see e.g. \cite{SNB,SN14} and references therein).}
\label{tab:parameter}
\end{table}
\vspace*{-0.2cm}
\section{Mass of the $ \bar D^*D(1^{++})$ molecule}
\vspace*{-0.2cm}
\subsubsection*{$\bullet$ $\tau$ and $t_c$ stabilities}
\begin{figure}[hbt] 
\centerline{\includegraphics[width=7.cm]{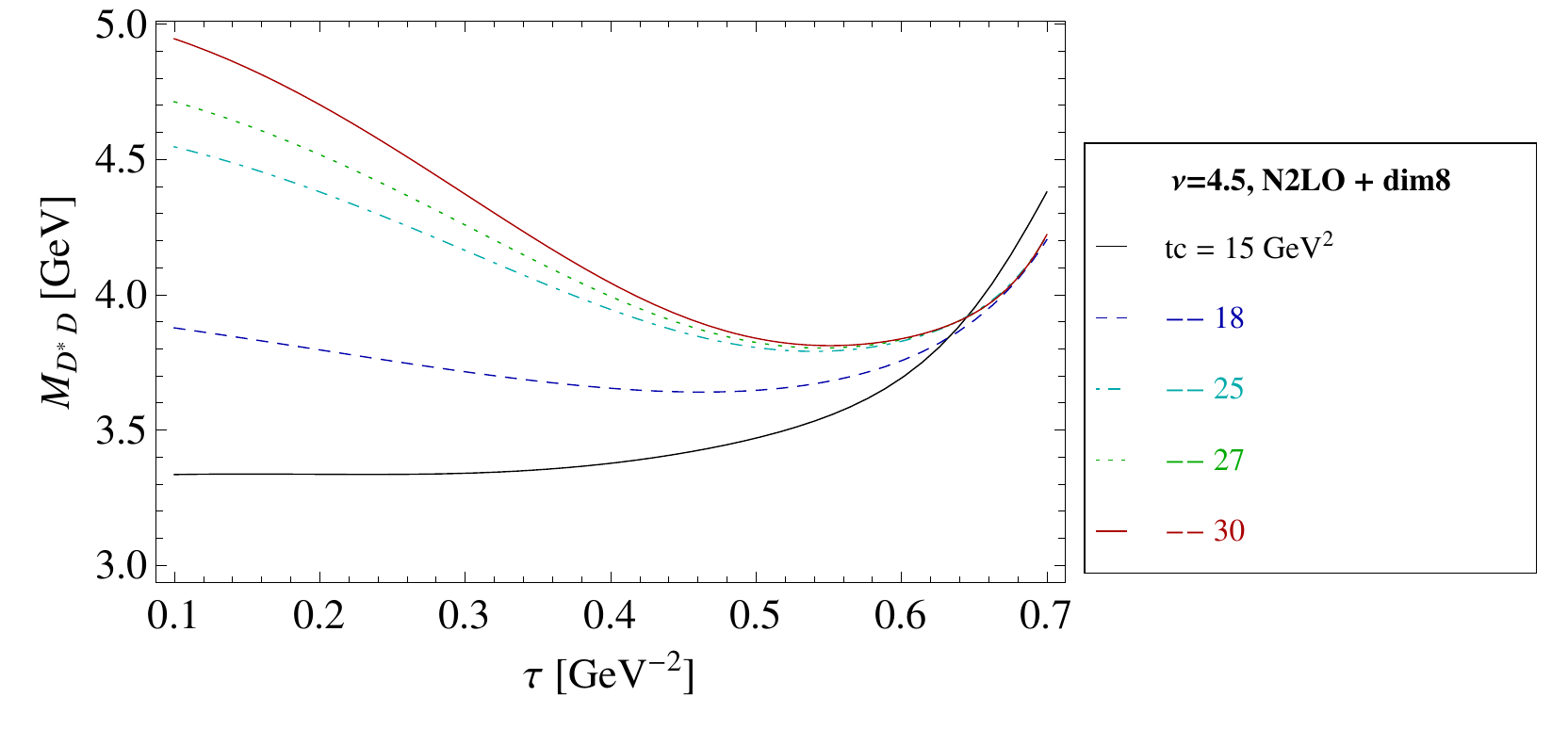}}
\caption{\scriptsize $\tau$-behavior of  $M_{D^*D}$ at N2LO for different values of $t_c$ and for $\nu$=4.5 GeV}
\label{fig1d} 
\end{figure}

\nin
 We study the behavior of the mass in term of LSR variable $\tau$ at different values of $t_c$ as shown in Fig.\ref{fig1d}. 
 We consider as final and conservative result the one corresponding to the beginning of the $\tau$ stability for $t_c$=18 GeV$^2$ until the one where $t_c$ stability is reached for $t_c\simeq$ 25 GeV$^2$.
\vspace*{-0.75cm}
\subsubsection*{$\bullet$ Convergence of the PT series} 
\begin{figure}[hbt] 
\centerline{\includegraphics[width=6.cm]{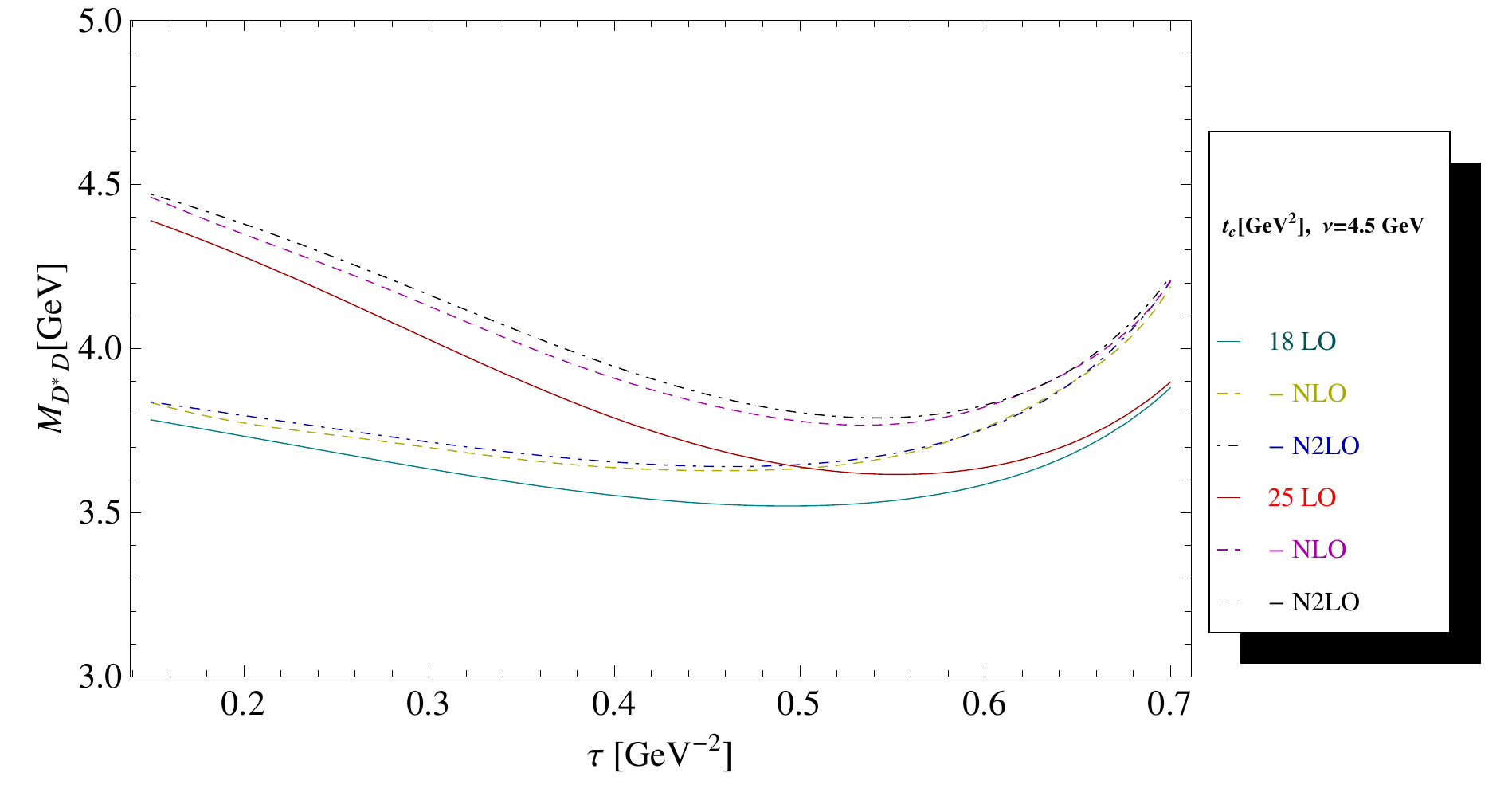}}
\caption{\scriptsize $\tau$-behavior of $M_{D^*D}$ for different values of $t_c$=18 and 25 GeV$^2$ and   $\nu$=4.5 GeV and for different truncation of the PT series}
\label{fig2d} 
\end{figure}

\nin
According to these analysis, we can notice that the $\tau$-stability begins at $t_c$=18 GeV$^2$ and the $t_c$-stability is reached from $t_c= 25$ GeV$^2$. Using these two extremal values of $t_c$, we study in Fig. {\ref{fig2d}} the convergence of the PT series for a given value of $\nu=4.5$ GeV.  We observe that from LO to NLO the mass increases by about +3.5$\%$ while  from NLO to N2LO, it only increases by +0.5$\%$. This result indicates a good convergence  of PT series which validates the LO result obtained in the literature when the  running quark mass is used \cite{RAPHAEL}.
\subsubsection*{$\bullet$ $\nu$-stability}
\begin{figure}[hbt] 
\centerline{\includegraphics[width=6.cm]{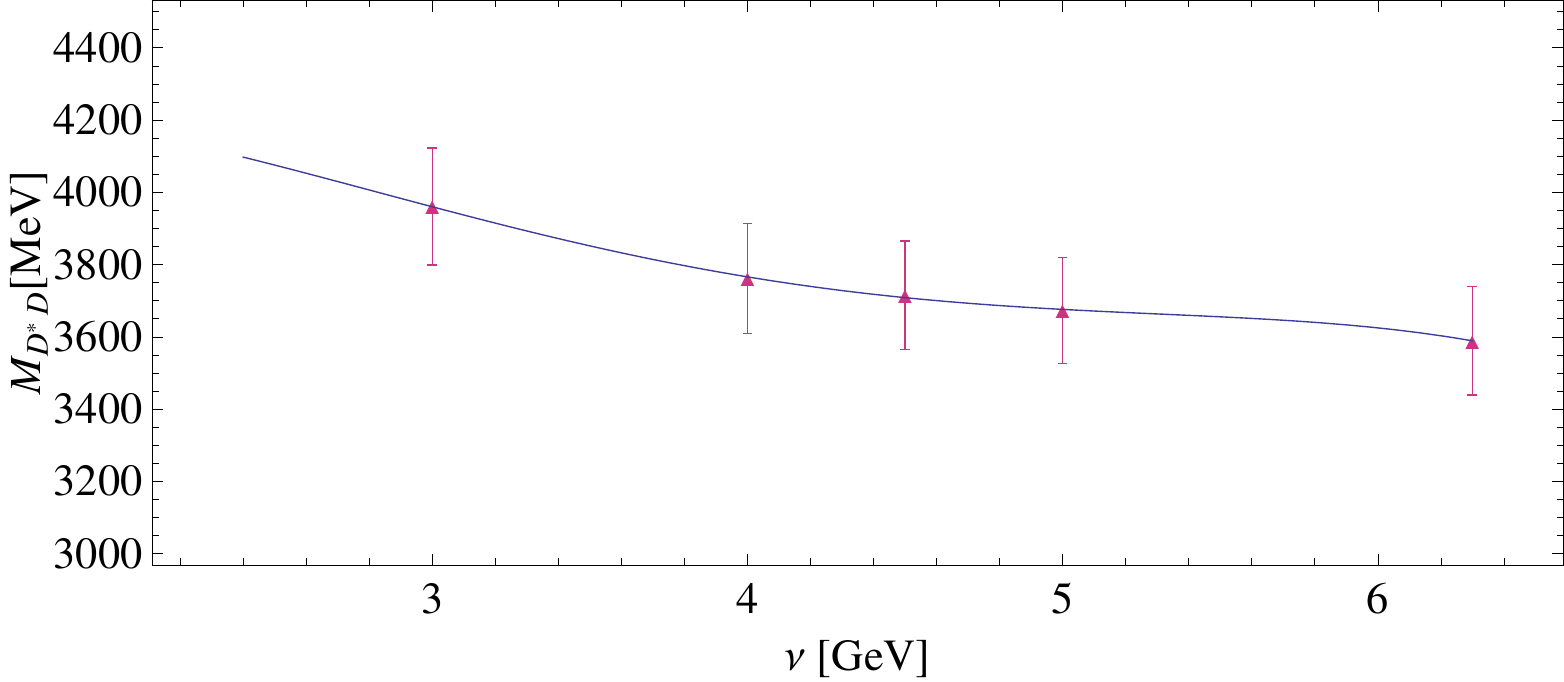}}
\caption{\scriptsize $\nu$-behaviour of $M_{D^*D}$  at N2LO}
\label{fig3d} 
\end{figure}

\nin
We improve our previous results by using different values of $\nu$ (Fig. {\ref{fig3d}}). Using the fact that 
the final result must be independent of the arbitrary parameter $\nu$, we consider as an optimal result the one at the inflexion point
for $\nu\simeq (4.0-4.5)$ GeV: 
\beq
 M_{D^*D} =3738(150)(23)~ \text{MeV}~,
\label{eq:md*d}
\eeq
where the second error comes from the localisation  of the inflexion point. This result agrees within the errors with the observed $Z_c(3900)$ candidate.
\vspace*{-0.5cm}
\section{Coupling of the $ \bar D^*D(1^{++})$ molecule}
\begin{figure}[hbt] 
\centerline{\includegraphics[width=6.cm]{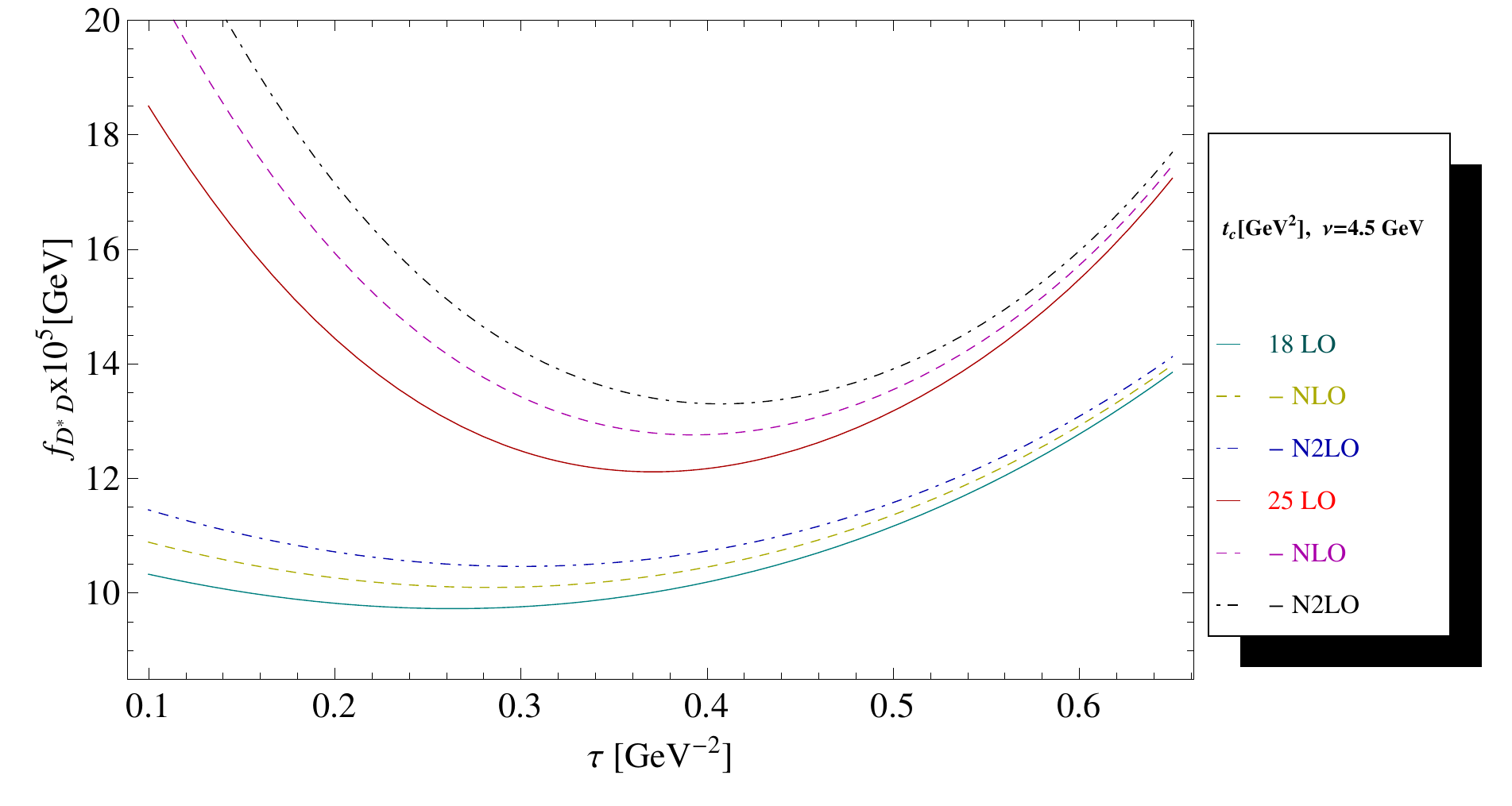}}
\caption{\scriptsize $\tau$-behavior of the running coupling $f_{D^*D}$ for $\nu=4.5$ GeV and for two extremal values of $t_c=18$ and 25 GeV$^2$.}
\label{fig1f} 
\end{figure}

\nin
We can do the same analysis to derive the decay constant $f_H$ defined in Eq. (\ref{eq:decay}). Noting that the bilinear pseudoscalar heavy-light current acquires an anomalous dimension, then the decay constant runs as:
\beq
f_H(\nu)=\hat f_H \ga {\rm Log}{\nu\over \Lambda}\dr^{2/-\beta_1}~,
\eeq
where $\hat f_H$ is a scale invariant coupling.
Taking the Laplace transform of the correlator, this definition will lead us to the expression of the running coupling in Eq. (\ref{coupling}). We show in Fig. \ref{fig1f} the $\tau$-behaviour of the running coupling $f_{D^*D}(\nu)$ for two extremal values of $t_c$ where $\tau$ and $t_c$ stabilities are reached. These values  are the same as in the mass determination. 
One can see in this figure that the $\alpha_s$ corrections to the LO term of PT series are still small though bigger than in the case of the mass determination from the ratio of sum rules. It is about +5.13$\%$ from LO to NLO  and +4.45$\%$ from NLO to N2LO.  
\begin{figure}[hbt] 
\centerline{\includegraphics[width=6.cm]{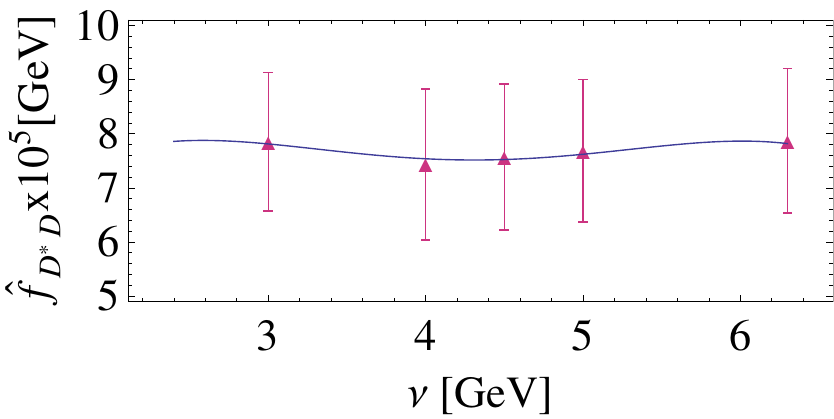}}
\caption{\scriptsize $\nu$-behavior of $f_{D^*D}$  at N2LO}
\label{fig2f} 
\end{figure} 
In the Fig. {\ref{fig2f}}, we show the $\nu$ behaviour of the invariant coupling $\hat f_{D^*D}$. Taking the optimal result at the minimum for $\nu\simeq 4$ GeV, we obtain in units of MeV:
\bea
\hat f_{D^*D}=(7.43\pm 1.40)\times 10^{-2}~{\rm MeV}~\lrar\nnb\\
 f_{D^*D}(\nu)=(11.57\pm 2.17)\times 10^{-2}~{\rm MeV}~,
\label{eq:fd*d}
\eea
which is comparable with the LO result  \cite{MATHEUS}: $f_{X_c}=(6.5\pm 1.1)\times 10^{-2}~{\rm MeV}$ appropriately normalized  of the X(3872). 
\section{ Mass and coupling of the $\bar B^*B(1^{++})$ molecule}
\begin{figure}[hbt] 
\centerline{\includegraphics[width=6.cm]{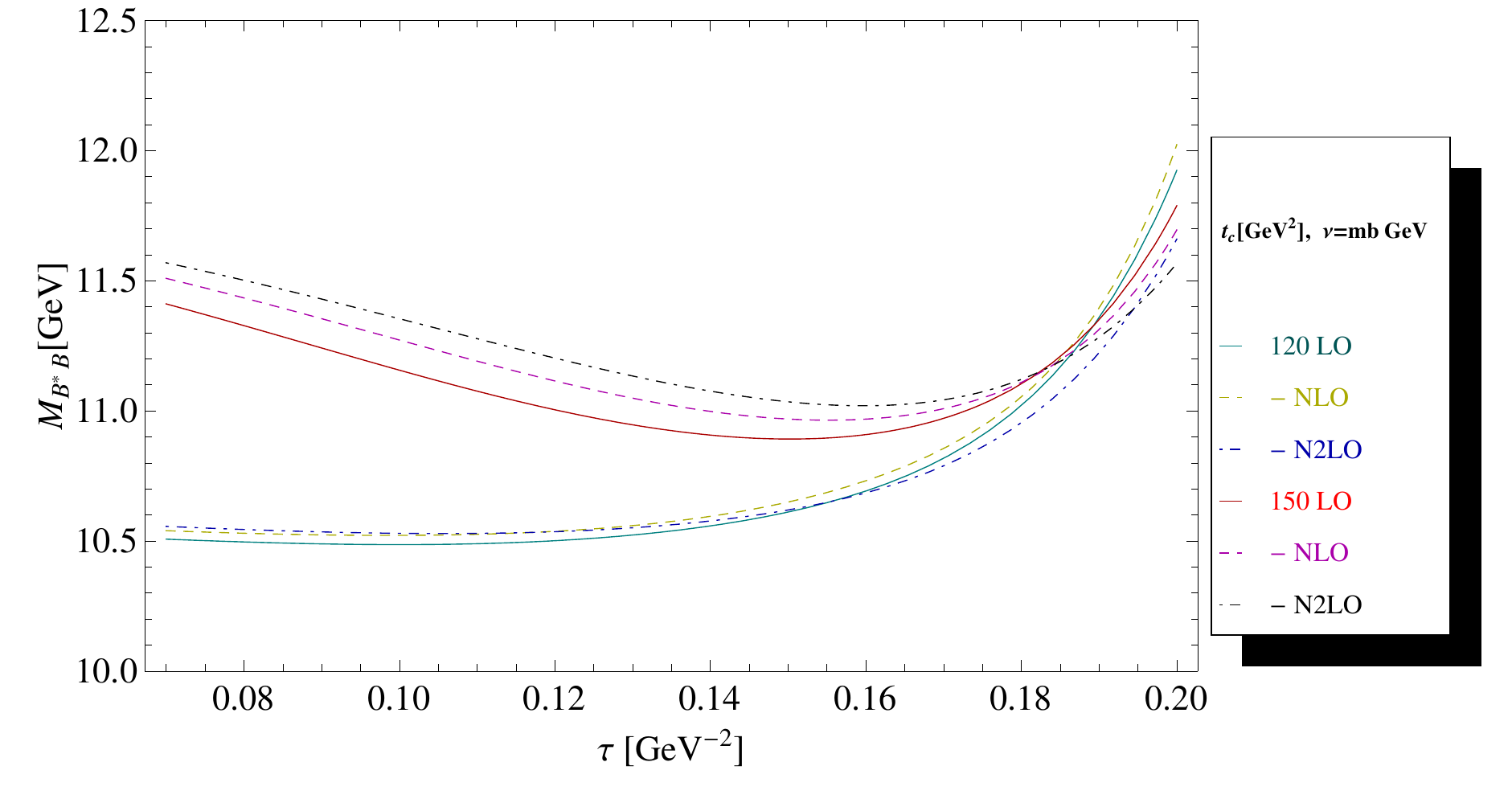}}
\caption{\scriptsize $\tau$-behavior of $M_{B^*B}$ with different values of $t_c$  for $\nu=\overline {m}_b(m_b)$ and for different
truncation of the PT series}
\label{fig1b} 
\end{figure}

\begin{figure}[hbt] 
\centerline{\includegraphics[width=6.cm]{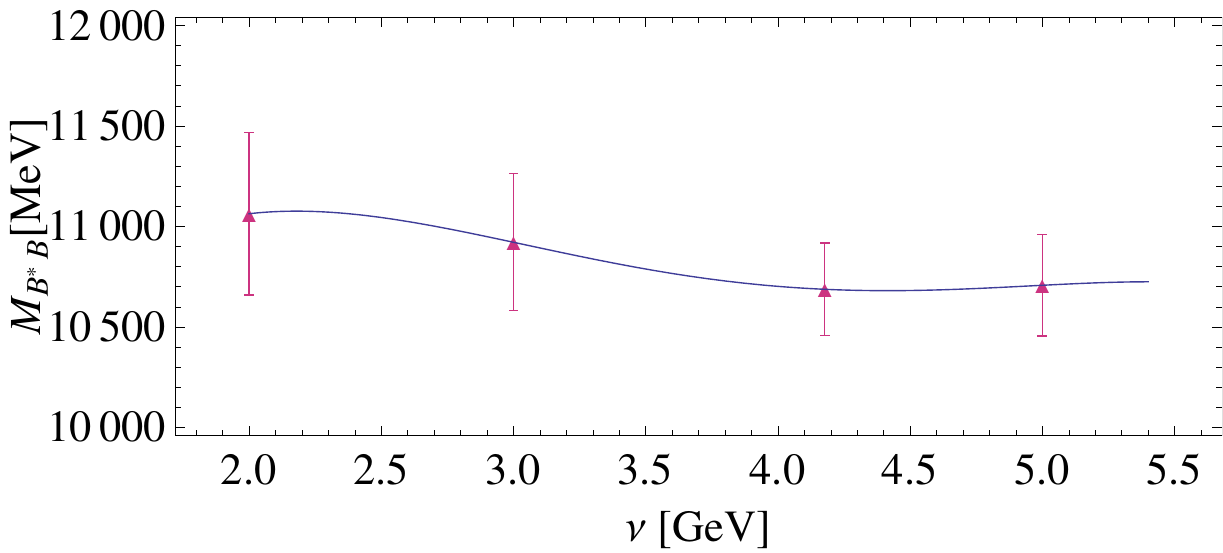}}
\caption{\scriptsize $\nu$-behavior of $M_{B^*B}$ mass  at N2LO}
\label{fig2b} 
\end{figure}

\begin{figure}[hbt] 
\centerline{\includegraphics[width=5.5cm]{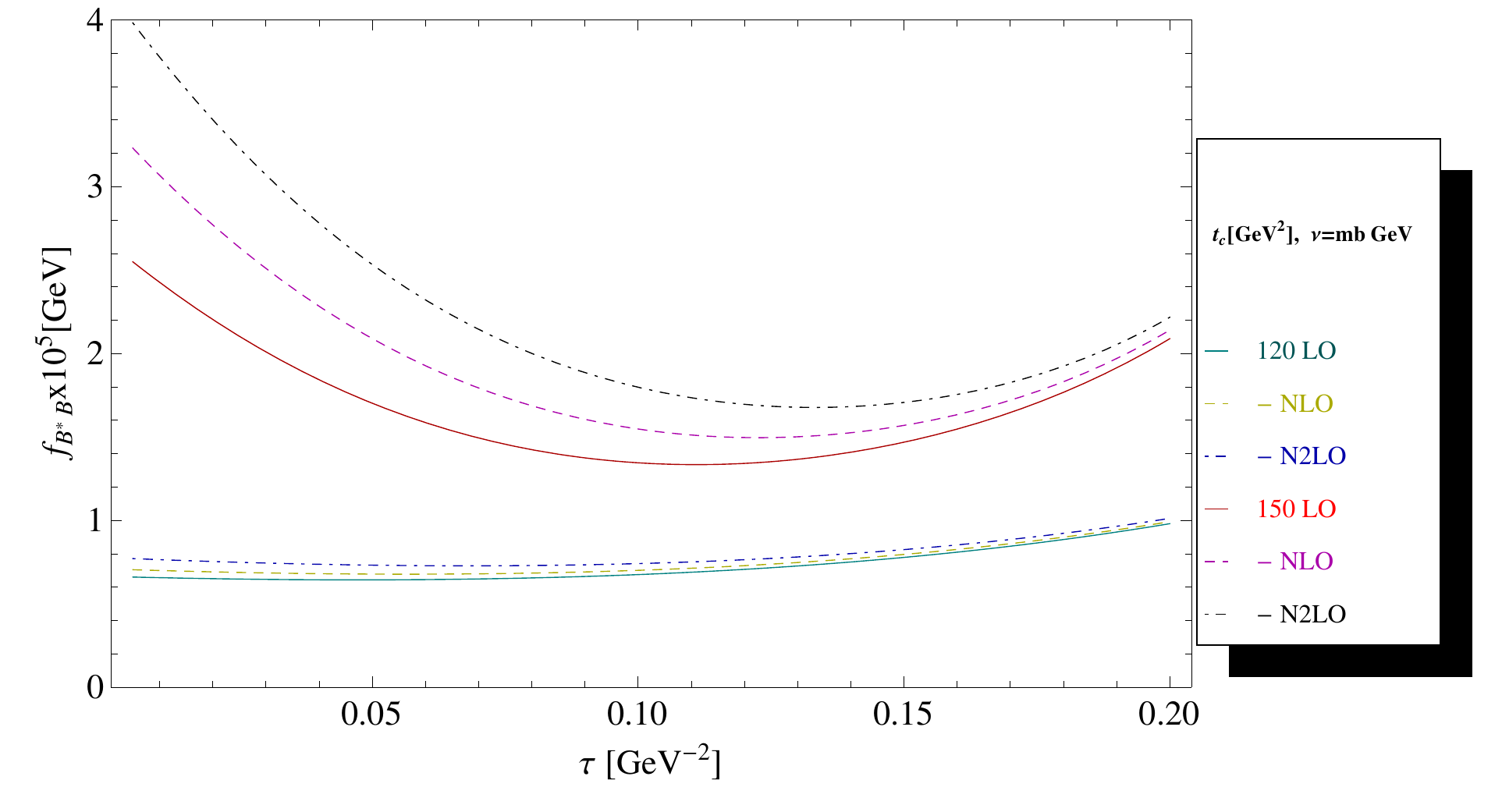}}
\caption{\scriptsize $\tau$-behavior of the running $ f_{B^*B}$ coupling for $\nu=\bar {m}_b(m_b)$, $t_c=120$ and 160 GeV$^2$ and
for different truncations of the PT series.}
\label{fig1bf} 
\end{figure}

\begin{figure}[hbt] 
\centerline{\includegraphics[width=6.cm]{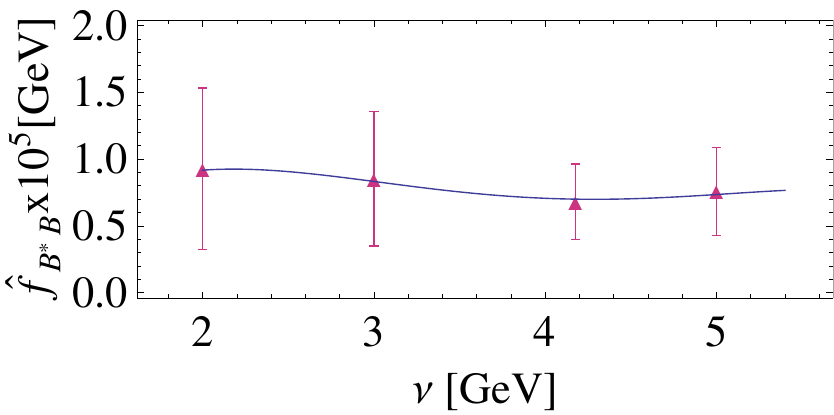}}
\caption{\scriptsize $\nu$-behavior of the invariant coupling $\hat f_{B^*B}$  at N2LO.}
\label{fig2bf} 
\end{figure}

\nin
We do the same analysis in the case of bottom channel. Fig. {\ref{fig1b}} shows the $\tau$-behavior of mass for $\nu=\overline {m}_b(m_b)$ and Fig. \ref{fig2b} shows its variation versus  $\nu$.
We have chosen two values of $t_c$ which correspond to the beginning of the $\tau$-stability ($t_c=$ 120 GeV$^2$) and to the beginning of $t_c$ stability ($t_c$=150 GeV$^2$). We observe a good convergence of PT series (increase of about 0.46$\%$ from LO to NLO and of about  0.35$\%$ from NLO to N2LO.
Considering the one at the minimum in $\nu=\overline{m}_b({m}_b)$ as the optimal result, we can deduce:
\bea
M_{B^*B}= 10687(232) \text{MeV}~,
\label{eq:mb*b}
\eea
where one can notice a good agreement with the observed $Z_b(10610)$ experimental candidate.
We show in the Fig.\,\ref{fig1bf} and Fig.\,\ref{fig2bf} the $\tau$ and $\nu$-behavior of the coupling for $\bar B^*B$. Like in the case of the charm channel, we will also have the same $t_c$ as in the determination of the mass.
\nin
Radiative corrections are more important here than in the case of ratio of moments as expected while the series is slowly convergent. From LO to NLO one has an increase of 10.1$\%$ and from NLO to N2LO an increase of about 9.4$\%$. The optimal result for the coupling is obtained at the minimum for $\nu=\overline{m}_b({m}_b)$:
 \bea
&&\hat f_{B^*B}=(0.69\pm 0.29) \times 10^{-2}~{\rm MeV}~~\lrar \nnb\\
&&f_{B^*B}(\nu)=(1.22\pm 0.51)  \times 10^{-2}~{\rm MeV}~,
\label{eq:fb*b}
\eea
again comparable with the LO result \cite{MATHEUS}: $f_{X_b}\simeq  10^{-2}$\,{\rm MeV} of the $X_b$ predicted at 10144(107) MeV.
\section{Conclusions}
\nin
We have presented improved predictions of QSSR for the masses and couplings of the $D^*D$ and $B^*B$ molecule states at N2LO  of PT series and including up to dimension 8 non-perturbative condensates. Our results given in Eqs.\,(\ref{eq:md*d}) and (\ref{eq:mb*b}) for the masses are in good agreement within the errors with the experimental candidates $Z_c(3900)$ and $Z_b(10610)$ suggesting that these new states may have large molecule components in their wave functions. However, if one extrapolate the result of Ref.\,\cite{SNPIVO} for $\bar B-B$ mixing,  where the breaking of the four-quark factorization is small (about 10$\%$ which should be explictily checked), one cannot exclude the four-quark assignement for these states. The couplings of these states to the corresponding interpolating currents are given in Eqs.\,(\ref{eq:fd*d})  and  (\ref{eq:fb*b}) and are comparable with the ones of the $X_c(3872)$ and $X_b(10144)$ predicted in \cite{MATHEUS}.
The extension of our analysis to some other molecule states is in progress.
\section*{Acknowledgments}
\nin
F.F. and A.R. would like to thank the CNRS for supporting the travel and living expenses and the LUPM-Montpellier for hospitality. We also thank R.M. Albuquerque for many helpful discussions.











\vspace*{-0.25cm}

\end{document}